\documentclass{elsart3p}
\usepackage{amsmath}
\usepackage{graphicx}

\renewenvironment{thebibliography}[1]{\begin{svbib}{#1}\tolerance=9999}{\end{svbib}}
\makeatletter
\let\oldps@copyright\ps@copyright
\renewcommand\ps@copyright{\oldps@copyright\renewcommand\@oddhead{\hfil\begin{minipage}[t]{3in}\begin{flushright}BNL-76794-2006-JA\\MUC-PUB-COOL\_THEORY-339\end{flushright}\end{minipage}}}
\makeatother

\DeclareMathOperator\re{Re}
\DeclareMathOperator\im{Im}
\newcommand\vv{\vec{v}}
\newcommand\zv{\vec{z}}
\newcommand\nablav{\vec{\nabla}}
\begin{document}
\begin{frontmatter}
  \title{Ionization Cooling in all Phase Space Planes with Various
  Absorber Shapes, Including Parallel-Faced Absorbers}
  \author{J. Scott Berg\thanksref{doe:bnl}}
  \date{25 July 2006}
  \thanks[doe:bnl]{Work Supported by the United States Department of Energy,
    Contract No.\ DE-AC02-98CH10886.}
  \ead{jsberg@bnl.gov}
  \ead[url]{http://pubweb.bnl.gov/people/jsberg/}
  \address{Brookhaven National Laboratory; Building 901A; P.O. Box 5000;
    Upton, NY  11973-5000}
  \begin{abstract}
    Ionization cooling in a straight beamline
    reduces the transverse emittance of a beam, and has little effect
    on the longitudinal emittance (generally, in fact, it increases
    the longitudinal emittance).  Once the beamline bends, the
    introduction of dispersion creates a coupling between the
    transverse and longitudinal planes.  If this coupling is handled
    properly, one can achieve cooling in all three phase space planes.
    This is usually done by placing a wedge-shaped absorber in a
    region where there is dispersion.  I will demonstrate using an
    eigenvalue analysis
    that there are other configurations of dispersion and absorber
    shape that will achieve ionization cooling in all phase space
    planes.  In particular, I will show that a one can even achieve
    cooling in all phase planes with a parallel-faced absorber in a
    dispersion-free region.  I will use perturbation theory to approximate the
    change in the cooling rates due to
    longitudinal-transverse coupling.  I will then describe how
    the cooling of longitudinal oscillations can be understood
    via the projection of the ``longitudinal'' eigenmodes into the
    transverse plane.
  \end{abstract}
  \journal{Nuclear Instruments and Methods A}
\end{frontmatter}

\section{Introduction}

Ionization cooling is a method for the rapid reduction of the
emittance of a beam by passing the beam through material
(hereafter called an
``absorber'')~\cite{atener19:534,atener31:40,sjpn12:223,pa14:75}.
Its primary
application has been to muon beams for a neutrino factory or muon
collider, but it has been contemplated for other applications
as well~\cite{ffag05:15,nima562:591,nima:2006:rubbia}.

In a
straight beamline, ionization cooling reduces only the transverse
emittance of a beam, generally having little effect on the
longitudinal emittance (in fact, generally making it somewhat worse).
It reduces the transverse emittance because the total momentum,
including the transverse, is reduced when when the beam passes
through material, but when the momentum is restored by an RF cavity,
the transverse momentum is left unchanged, and thus reduced from
its value before the absorber.  It is desirable, especially for
a muon collider, to reduce the longitudinal emittance as well.

\begin{figure}
  \includegraphics[width=\linewidth]{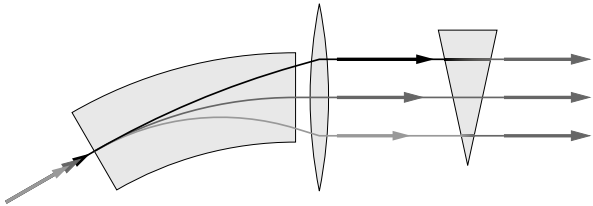}
  \caption{Dispersion generated by a bending magnet (left) means that
    particles with different energies have different positions.  The
    wedge-shaped absorber (right) then reduces the energies of higher
    energy particles more than it reduces the energies of lower energy
    particles.}
  \label{fig:emitexch}
\end{figure}There is a well-known method for accomplishing this, often
referred to as ``emittance exchange''~\cite{pa14:75,ichea12:481,ichea12:485}.
This is generally accomplished by using
a wedge-shaped absorber in a region with non-zero
dispersion, as shown in Fig.~\ref{fig:emitexch}.  Particles with higher
energy will pass through more material than particles with
lower energy as a result of the dispersion.  Thus, the energy
spread in the beam will be reduced.  There is a cost to this,
however.  Imagine that all the energy spread were removed by this.  Then
there would be no way of removing the spread in transverse
position that was introduced by the dispersion.  Thus, reduction
in longitudinal emittance comes at a cost: the horizontal
emittance is increased.  If done properly, the cooling accomplished
transversely by an absorber can be shared between the longitudinal
and transverse planes, giving emittance reduction in all
phase space planes.

This paper will demonstrate that achieving cooling in the longitudinal
plane can be accomplished by a more general class of methods than just
having dispersion in transverse position at a wedge-shaped absorber.
It will demonstrate that cooling in all planes happens when there is
an appropriate coupling between the longitudinal and transverse planes.
The paper will explore multiple ways of generating that coupling
and compare their effectiveness.

I will first describe a simplified lattice that will be used in this
paper.  I will then ascertain whether the lattice is cooling in
all phase space planes by examining the eigenvalues for the linear
matrix representing that lattice.  I will compute the eigenvalues
approximately using perturbation theory, where the small parameter
in the expansion is the degree of coupling between the longitudinal
and transverse planes.  I will study coupling that is generated by dispersion
and an appropriately shaped absorber, and coupling generated by
dispersion in the RF cavity.  I will then look at the eigenvalues 
for cases when the perturbation expansion might not be accurate.
Finally, I will develop a more physically intuitive understanding
of how transverse cooling can reduce the longitudinal emittance
by examining the motion in the coupled system using the eigenvectors
of the matrix that represents the lattice, which will give a picture
of the coupled motion when it is projected into the longitudinal and
transverse planes.

This paper does not address other issues of cooling lattice design,
in particular the final equilibrium emittance which results because
of multiple scattering and energy straggling.  The paper is intended
only to present a wider range of ideas on how cooling can be accomplished
in all phase space planes, not to present a full generalized theory of
ionization cooling or a working cooling lattice design.

\section{Lattice}

The cooling lattice will consist of a sequence of identical cells with
four sections:
\begin{enumerate}
\item An absorber, which reduces the energy of the particles
\item An RF cavity, which restores the energy lost in the absorber
  and provides longitudinal focusing
\item A group of magnets transporting the beam from the absorber
  to the next RF cavity
\item A group of magnets transporting the beam from the cavity
  to the next absorber
\end{enumerate}
The details of what is in the ``group of magnets'' will not be important
for this discussion.

There is a planar reference curve which defines the coordinate system
that we are using.  The vertical coordinate $y$ is perpendicular to
the plane in which the curve lies.  The horizontal coordinate $x$ is
perpendicular to the vertical and to the tangent to the curve, and
stays on the same side of the curve.  The arc length along the curve,
$s$, is the independent variable for the equations of motion.  For
the examples in this paper, the vertical magnetic field at the
reference curve $B_0(s)$ will be such that a particle with a reference
energy $E_{\text{r}}(s)$ which starts out on the curve and moving tangent to
the curve continues to do so.  This particle will be referred to as
the reference particle.  Note that the reference energy depends on the
position in the lattice, due to the absorber and the RF cavity.

The magnetic field will be midplane symmetric, meaning that the
vertical field is symmetric under $y\to-y$, and the other magnetic fields are
antisymmetric under $y\to-y$.  There are two consequences of this.
The first is that this is not a solenoid focused lattice, which is
typical for an ionization cooling lattice, but a quadrupole-focused
lattice.  The second is that the vertical motion is, to lowest order,
decoupled from the horizontal and longitudinal motions.  This second
reason is the fundamental purpose in choosing a midplane symmetric
lattice.  The qualitative results found here will continue to hold
for a solenoid-focused lattice.

The reference particle will lose an energy $\Delta E$ in the absorber
and will gain back that same amount energy in the RF cavity.
The energies in the magnets between these elements will be
$E_\pm=E_0\pm\Delta E/2$, the corresponding total momenta are
$p_{\pm}=\sqrt{E_\pm^2/c^2-(mc)^2}$, and
$p_0=\sqrt{E_0^2/c^2-(mc)^2}$.  In these expressions $m$ is the
particle mass and $c$ is the speed of light.

\subsection{Matrices for Lattice Sections}
I want to determine whether for small deviations from the particle with
coordinates zero, the amplitude of oscillations is growing or reducing.
Thus, this paper will only keep results to first order in these deviations,
and will use matrices to represent how these deviations change while
propagating through the lattice.  The lattice is a
sequence of identical cells, and the eigenvalues for the matrix for an
entire cell will determine the characteristics of the motion.

I will denote the absorber location by a subscript ``a,'' and the
cavity location by a subscript ``c.''  The matrix describing the
motion through the magnets from the absorber to the cavity is
$\mathcal{M}_{\text{ca}}$, and the matrix from the cavity to the
absorber is $\mathcal{M}_{\text{ac}}$.  I will assume a kind of
reflection symmetry at the absorber and cavity such that these
matrices can be written as
\begin{align}
  \mathcal{M}_{\text{ac}} &= \mathcal{D}_{\text{a}}\mathcal{B}_{\text{a}}
  \mathcal{R}(\mu_-,R_{56}^-)
  \mathcal{B}_{\text{c}}^{-1}\mathcal{D}_{\text{c}}^{-1}\\
  \mathcal{M}_{\text{ca}} &= \mathcal{D}_{\text{c}}\mathcal{B}_{\text{c}}
  \mathcal{R}(\mu_+,R_{56}^+)
  \mathcal{B}_{\text{a}}^{-1}\mathcal{D}_{\text{a}}^{-1},
\end{align}
where
\begin{align}
  \mathcal{B}_i &=
  \begin{bmatrix}
    b_i^{1/2}&0&0&0\\0&b_i^{-1/2}&0&0\\0&0&1&0\\0&0&0&1
  \end{bmatrix}\\
  \mathcal{R}(\mu,R_{56}) &=
  \begin{bmatrix}
    \cos\mu&\sin\mu&0&0\\
    -\sin\mu&\cos\mu&0&0\\
    0&0&1&R_{56}\\
    0&0&0&1
  \end{bmatrix}\\
  \mathcal{D}_i &=
  \begin{bmatrix}
    1&0&0&d_i^x\\
    0&1&0&d_i^p\\
    d_i^p&-d_i^x&1&0\\
    0&0&0&1
  \end{bmatrix}.
\end{align}
These quantities are related to the usual accelerator quantities
by $b_i=\beta_x(s_i)/p_{\text{r}}(s_i)$,
$d_i^x=D(s_i)/[\beta_{\text{r}}(s_i)p_{\text{r}}(s_i)c]$,
and $d_i^p=D'(s_i)/[\beta_{\text{r}}(s_i)c]$, where
$p_{\text{r}}(s)=\sqrt{E^2_{\text{r}}(s)/c^2-(mc)^2}$,
$\beta_{\text{r}}(s)=p_{\text{r}}(s)c/E_{\text{r}}(s)$, $\beta_x(s)$
is the Courant-Snyder beta function, and $D(s)$ is the dispersion.
Note that since $p_{\text{r}}(s_{\text{a}})=p_\pm$ depending on which
side of the absorber one is on, the beta functions on each side
of the absorber are actually slightly different.  The same type of
difference exists
for the cavity and the dispersion.

Assuming that the only effect of the RF cavity and absorber is to
shift the energy of all particles by $\pm\Delta E$, the matrix
for the full cell from the absorber back to the next absorber
is
\begin{equation}
  \mathcal{D}_{\text{a}}\mathcal{B}_{\text{a}}
  \mathcal{R}(\mu_-+\mu_+,R_{56}^-+R_{56}^+)
  \mathcal{B}_{\text{a}}^{-1}\mathcal{D}_{\text{a}}^{-1}
\end{equation}
Thus, the transverse phase advance per cell is $\mu_-+\mu_+$, and
$R_{56}^-+R_{56}^+$ is approximately $\eta_cTE/(pc)^2$, where
$\eta_c$ is the frequency slip factor, and $T$ is the time for
the reference particle to go through one cell.  The momentum $p$
and energy $E$ are intentionally left ambiguous: $\eta_c$ is
only defined for a fixed energy.

The RF cavity is described by a matrix
\begin{equation}
  \mathcal{V}=
  \begin{bmatrix}
    1&0&0&0\\0&1&0&0\\0&0&1&0\\
    0&0&\omega V\cos\phi&1
  \end{bmatrix},
\end{equation}
where $V$ is the maximum energy gain in the cavity, the RF phase is $\phi$,
where $\phi=0$ is the phase for zero energy gain, and $\omega$ is
$2\pi$ times the RF frequency.

The absorber reduces the energy of the particles, maintaining their
direction.  The horizontal position will thus be the same as it would be
if the absorber were a drift.  The time to traverse the absorber
is well approximated by assuming that the momentum is $p_+$ to the center
of the absorber, then $p_-$ thereafter.  Thus, it is a good approximation
to treat the absorber as a thin element, with surrounding drifts
absorbed into $\mathcal{M}_{\text{ac}}$ and $\mathcal{M}_{\text{ca}}$.

Since the particle direction is maintained, the transverse momentum
will be reduced in the absorber by the same factor that the total
momentum is reduced, $\kappa_x=p_-/p_+$.  Particles whose energy
differs from the reference energy will receive a slightly different
energy loss than the reference particle, since the energy loss per
unit length in the absorber, $dE/dx$, depends on energy.  To linear order,
the energy loss of a particle with energy $E$ just before the absorber is
\begin{multline}
  \Delta E+
  \Delta E\left(\dfrac{dE}{dx}\right)^{-1}
  \dfrac{d}{dE}\left(\dfrac{dE}{dx}\right)(E-E_+)\\
  \mbox{}=\Delta E+(1-\kappa_z)(E-E_+)
\end{multline}

\begin{figure}
  \includegraphics[width=\linewidth]{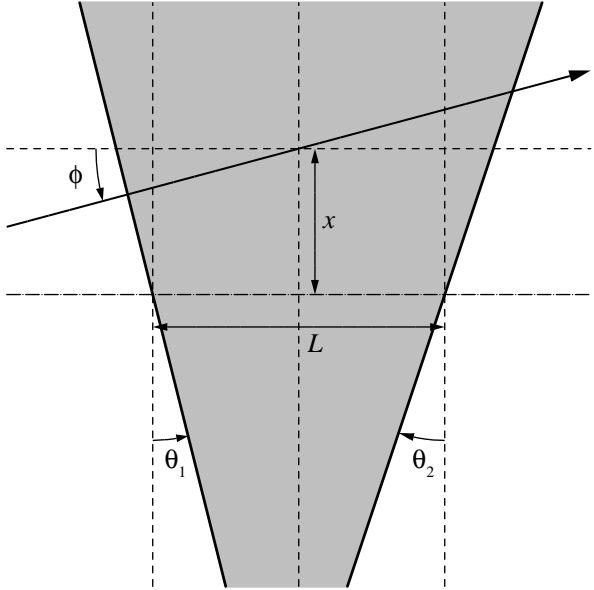}
  \caption{Geometry of the absorber.  $\sin\phi=p_x/p_+$, where $p_x$
    is the horizontal momentum before the absorber.}
  \label{fig:wedge}
\end{figure}
Furthermore, if the faces of the absorber are not
perpendicular to the reference orbit, there may be an energy loss
which depends on the horizontal position or angle (again,
I'm ignoring the vertical motion).
Using the geometry from Fig.~\ref{fig:wedge},
the change in the energy is $a_+x+a_-Lp_x/(2p_+)$~\cite{pac03:2210}, where
\begin{equation}
  a_\pm = -\dfrac{dE}{dx}(\tan\theta_2\pm\tan\theta_1).
\end{equation}
These define the matrix for the absorber, which I will denote
$\mathcal{A}$:
\begin{equation}
  \mathcal{A} =
  \begin{bmatrix}
    1&0&0&0\\
    0&\kappa_x&0&0\\
    0&0&1&0\\
    a_+&a_-L/(2p_+)&0&\kappa_z
  \end{bmatrix}.
\end{equation}

\section{Eigenvalue Analysis}

The transfer matrix for the full lattice,including the absorbers and the
RF cavities, is
\begin{equation}
  \mathcal{A}\mathcal{M}_{\text{ac}}\mathcal{V}\mathcal{M}_{\text{ca}}.
\end{equation}
I will be analyzing the eigenvalues of this matrix to determine if I
am achieving cooling in both planes.  First of all, it is clear that the
product of the eigenvalues is $\kappa_x\kappa_z$, and that the
eigenvalues will come in complex conjugate pairs.

Begin with a basic lattice, with no dispersion, and with
the absorber face angle $\theta_1=\theta_2=0$.
The characteristic polynomial for the matrix for the cell is
\begin{multline}
  (\lambda^2-\lambda(1+\kappa_z+(R_{56}^-+\kappa_zR_{56}^+)\omega V\cos\phi)
  +\kappa_z)\\
  (\lambda^2-\lambda(1+\kappa_x)\cos(\mu_++\mu_-)+\kappa_x).
  \label{eq:basepoly}
\end{multline}
The roots of the characteristic polynomial are the eigenvalues.

From this polynomial, we can see that the eigenvalues will be of the
form
\begin{align}
  \sqrt{\kappa_x}e^{\pm i\mu_x}&&
  \sqrt{\kappa_z}e^{\pm i\mu_z}.
\end{align}
It turns out that the eigenvalues remain the same if $d_{\text{a}}^x$ and
$d_{\text{a}}^p$ are nonzero, as long as one minor correction is applied:
in $\mathcal{M}_{\text{ac}}$, $d_{\text{a}}^x$ should be replaced with
$d_{\text{a}}^x\sqrt{\kappa_x}$, and in $\mathcal{M}_{\text{ca}}$, 
$d_{\text{a}}^x$ should be replaced with $d_{\text{a}}^x/\sqrt{\kappa_x}$.

Thus, $\sqrt{\kappa_x}$ is the magnitude of the transverse eigenvalue,
and $\sqrt{\kappa_z}$ is the magnitude of the longitudinal eigenvalue.
Clearly $\kappa_x<1$; unfortunately, in most cases of interest for
cooling, $\kappa_z>1$.  However, the product $\kappa_x\kappa_z<1$.
$\kappa_z>1$ because for low energies, $(d/dE)(dE/dx)<0$.  While
$(d/dE)(dE/dx)$ becomes positive for higher energies, the relative
energy loss for a given absorber
is less at higher energies, making $\kappa_x$ closer to 1,
and the product $\kappa_x\kappa_z$ closer to 1.

Since the product $\kappa_x\kappa_z<1$, if one could generate coupling
between the horizontal and vertical motion, it would be possible to
reduce the magnitude of the ``longitudinal'' eigenvalues (they begin
to lose their identity with coupling), at the cost of raising
the magnitude of the ``transverse'' eigenvalues.  One could do so in
a way that made the magnitudes of
both sets of eigenvalues less than 1, and thus gave
cooling in all phase space planes.

If the characteristic polynomial for a matrix is
Eq.~(\ref{eq:basepoly}) plus some additional polynomial $f(\lambda)$,
then one can compute the change in the eigenvalues to lowest order.
The lowest-order change in the magnitude of
$\lambda_x=\sqrt{\kappa_x}e^{i\mu_x}$ is
\begin{equation}
  -\dfrac{\csc\mu_x}{2\sqrt{\kappa_x}}
  \im\left\{\dfrac{e^{-i\mu_x}f(\lambda_x)}
    {(\lambda_x-\lambda_z)(\lambda_x-\lambda^*_z)}\right\}
\end{equation}
and the lowest-order change in the magnitude of
$\lambda_z=\sqrt{\kappa_z}e^{i\mu_z}$ is
\begin{equation}
  -\dfrac{\csc\mu_z}{2\sqrt{\kappa_z}}
  \im\left\{\dfrac{e^{-i\mu_z}f(\lambda_z)}
    {(\lambda_z-\lambda_x)(\lambda_z-\lambda^*_x)}\right\}.
\end{equation}
Since a change in the magnitude of $\lambda_z$ leads to a corresponding
change in the magnitude of $\lambda_x$, one only need examine the
change in one eigenvalue.

Note that when $\lambda_x$ is close to $\lambda_z$ or $\lambda_z^*$,
there will be a large change in the eigenvalues.  This corresponds to
a linear coupling resonance between the longitudinal and horizontal.
For our perturbation analysis, we will assume that we are sufficiently
far from that resonance condition, but this will be of interest later
in the paper nonetheless.

Since $f$ is the difference between the characteristic polynomial for
the full system and Eq.~(\ref{eq:basepoly}), we know something more
about its properties.  First of all, it is a third order polynomial,
since all characteristic polynomials of a $4\times4$ matrix are
fourth order polynomials with leading order term $\lambda^4$.  Secondly,
the constant term in the characteristic polynomial is the determinant,
and the determinant of the matrix is $\kappa_x\kappa_z$ in all cases.
Thus, $f$ will have no constant term.  We can thus write $f(\lambda)$ as
$f_3\lambda^3+f_2\lambda^2+f_1\lambda$.  

I will examine the change in the magnitude of $\lambda_z$.
Using the polynomial expression
for $f(\lambda)$, it is
\begin{equation}
  \dfrac{
    \begin{aligned}
      &2\sqrt{\kappa_x\kappa_z}(\kappa_zf_3-f_1)\cos\mu_x\\
      &\mspace{2mu}\mbox{}-2\kappa_z(\kappa_xf_3-f_1)\cos\mu_z%\\
      %&\mspace{195mu}\mbox{}
      +\sqrt{\kappa_z}(\kappa_z-\kappa_x)f_2
    \end{aligned}
  }
  {
    \begin{aligned}
      &2[\kappa_z+\kappa_x-2\sqrt{\kappa_x\kappa_z}\cos(\mu_x-\mu_z)]\\
      &\mspace{80mu}
      [\kappa_z+\kappa_x-2\sqrt{\kappa_x\kappa_z}\cos(\mu_x+\mu_z)]
    \end{aligned}
  }.
  \label{eq:fcoef}
\end{equation}

There are two methods for generating coupling between the phase space
planes. The most common method is to create an absorber with nonzero
values for the angles $\theta_1$ and $\theta_2$ (in most cases with
$\theta_1=\theta_2$, thus referred to as a ``wedge'').  Another method
is to generate the coupling by having dispersion in the RF cavity,
which does not require any change in the shape of the absorber.

\subsection{Rotated Absorber Faces}

The additional terms in the characteristic polynomial when
$\theta_0$ or $\theta_1$ are nonzero but the dispersion at the cavity
is zero are
\begin{multline}
  \dfrac{a_+d^x_{\text{a}}}{\sqrt{\kappa_x}}[
  \lambda^3(\cos\mu_{x0}-\kappa_xs_+)\\
  \mbox{}+\lambda^2\cos\mu_{x0}(\kappa_x^2s_+-s_-)\\
  \shoveright{\mbox{}+\lambda\kappa_x(s_--\kappa_x\cos\mu_{x0})]}\\
  \mbox{}+a_+d^p_{\text{a}}b_{\text{a}}\sin\mu_{x0}
  [\lambda^3-\lambda^2(\kappa_xs_++s_-)+\lambda\kappa_x]\\
  \mbox{}-\dfrac{a_-Ld_{\text{a}}^x}{2p_+\sqrt{\kappa_x}b_{\text{a}}}
  \sin\mu_{x0}
  [\lambda^3-\lambda^2(\kappa_xs_++s_-)+\lambda\kappa_x]\\
  \shoveleft{
    \mbox{}+\dfrac{a_-Ld^p_{\text{a}}}{2p_+}[\lambda^3(\cos\mu_{x0}-s_+)
  }\\
  \mbox{}+\lambda^2\cos\mu_{x0}(s_+-s_-)\\
  \mbox{}+\lambda(s_--\cos\mu_{x0})],
\end{multline}
where $\mu_{x0}=\mu_++\mu_-$ and $s_\pm=1+R_{56}^\pm\omega V\cos\phi$.
Note that from Eq.~(\ref{eq:basepoly}),
$\kappa_zs_++s_-=2\sqrt{\kappa_z}\cos\mu_z$.  The numerator of
Eq.~(\ref{eq:fcoef}) for this case is $\sqrt{\kappa_z}$ times
\begin{multline}
  \dfrac{a_+d_{\text{a}}^x}{\sqrt{\kappa_x}}
  [s_-+\kappa_xs_+-(1+\kappa_x)\cos\mu_{x0}]\\
  \shoveright{[\kappa_x(s_-+s_+\kappa_z)-(\kappa_x^2+\kappa_z)\cos\mu_{x0}]}\\
  \shoveleft{\mbox{}-a_+d_{\text{a}}^pb_{\text{a}}
    (\kappa_z-\kappa_x)\sin\mu_{x0}}\\
  \shoveright{[s_-+\kappa_xs_+-(1+\kappa_x)\cos\mu_{x0}]}\\
  \shoveleft{\mbox{}+\dfrac{a_-Ld_{\text{a}}^x}
    {2p_+\sqrt{\kappa_x}b_{\text{a}}}
    (\kappa_z-\kappa_x)\sin\mu_{x0}}\\
  \shoveright{[s_-+\kappa_xs_+-(1+\kappa_x)\cos\mu_{x0}]}\\
  \shoveleft{\mbox{}+\dfrac{a_-Ld_{\text{a}}^p}{2p_+}
    [s_-+\kappa_xs_+-(1+\kappa_x)\cos\mu_{x0}]}\\
  [s_-+s_+\kappa_z-(1+\kappa_z)\cos\mu_{x0}].
  \label{eq:rotabs}
\end{multline}

Since $\kappa_x\approx\kappa_z\approx1$ in most cases, the quantity
$\kappa_z-\kappa_x$ can be treated as small.  Thus, the second and third terms in Eq.~(\ref{eq:rotabs}) can be neglected, as long as one is far
from the linear synchro-betatron resonance.  Equation~(\ref{eq:rotabs})
indicates that if one has a dispersion in position but not momentum at
the absorber, the absorber should have $\theta_1=\theta_2$ (i.e.,
a wedge shape), and $\theta_1$ positive if the dispersion is positive.
This is the usual method of ``emittance exchange.''

If, on the other hand, there is no dispersion in position but there
is dispersion in horizontal momentum, one should instead use a rotated slab
(i.e., $\theta_1=-\theta_2$).  Particles with larger momentum will
have an angle in one direction; if the absorber is rotated so there
is more material along that direction, there will be a greater energy
loss for those particles.

One might ask whether it is easier to construct a lattice that makes the
first term large than it is to construct a lattice that makes the
fourth term large.  In fact, either term can be made large, and which
one is used
depends on the desired properties of the lattice.  Take, for instance,
the RFOFO cooling ring described in~\cite{prstab8:061003}.
$(dE/dx)d_{\text{a}}^x$ for that lattice is about 0.027.  If one
finds the largest value of the momentum dispersion, then
$(dE/dx)Ld_{\text{a}}^p/(2p_+)$ is around 0.002.  While it may seem that
the former is significantly larger than the latter, the fact is that
the lattice was designed with a ring shape, which tends to generate
position dispersion rather than momentum dispersion.  One can
construct lattices where $(dE/dx)Ld_{\text{a}}^p/(2p_+)$ is larger, especially
when the beta function at the absorber is small, by having one half
of the lattice cell bending in one direction and and the other half bending
in the reverse direction.

\subsection{Dispersion in the RF Cavity}

If instead, $\theta_0=\theta_1=0$, but we have dispersion at the
cavity, there is still coupling between the longitudinal and transverse
planes, in this case generated by the dispersion at the RF cavity.  The
additional terms in the characteristic polynomial are
\begin{multline}
  \dfrac{\omega V\cos\phi}{2b_{\text{c}}}\lambda(\lambda-1)(\lambda-\kappa_z)\\
  \{[(d_{\text{c}}^x)^2+b_{\text{c}}^2(d_{\text{c}}^p)^2](\kappa_x+1)
  \sin(\mu_++\mu_-)\\
  \mbox{}-[(d_{\text{c}}^x)^2-b_{\text{c}}^2(d_{\text{c}}^p)^2](\kappa_x-1)
  \sin(\mu_+-\mu_-)\\
  \mbox{}-2b_{\text{c}}d_{\text{c}}^xd_{\text{c}}^p(\kappa_x-1)
  \cos(\mu_+-\mu_-)\}
\end{multline}
The numerator of Eq.~(\ref{eq:fcoef}) is then
\begin{multline}
  \dfrac{\omega V\cos\phi}{2b_{\text{c}}}(\kappa_z-\kappa_x)
  [2\kappa_z\cos\mu_z-\sqrt{\kappa_z}(1+\kappa_z)]\\
  \{[(d_{\text{c}}^x)^2+b_{\text{c}}^2(d_{\text{c}}^p)^2](\kappa_x+1)
  \sin(\mu_++\mu_-)\\
  \mbox{}-[(d_{\text{c}}^x)^2-b_{\text{c}}^2(d_{\text{c}}^p)^2](\kappa_x-1)
  \sin(\mu_+-\mu_-)\\
  \mbox{}-2b_{\text{c}}d_{\text{c}}^xd_{\text{c}}^p(\kappa_x-1)
  \cos(\mu_+-\mu_-)\}
  \label{eq:dispcav}
\end{multline}
For the RFOFO ring in~\cite{prstab8:061003}, the quantity
\begin{equation}
  \dfrac{\omega V\cos\phi}{2b_{\text{c}}}(d_{\text{c}})^2
\end{equation}
is around
$1.7\times10^{-3}$.  The product of the next two factors in
Eq.~(\ref{eq:dispcav}) is around $-0.02$.  Thus, compared to an
absorber with substantial rotations in its faces, the strength
of the coupling generated by dispersion in the RF cavities is
relatively weak.  The RFOFO ring was not necessarily designed
for maximizing coupling in this way, but it is unlikely to be
possible to increase the dispersion sufficiently while keeping
a reasonably small beta function in the lattice.  One may
be able to increase the RF voltage somewhat, but certainly not
enough to change the overall result.

\subsection{Running Closer to Resonance}

The perturbation theory analysis assumed that one was far from
the point where any two of the eigenvalues of the matrix are
close to each other.  If they do become close, then the estimates
above are not valid, and the eigenvalues should be computed directly.

\begin{figure}
  \includegraphics[width=\linewidth]{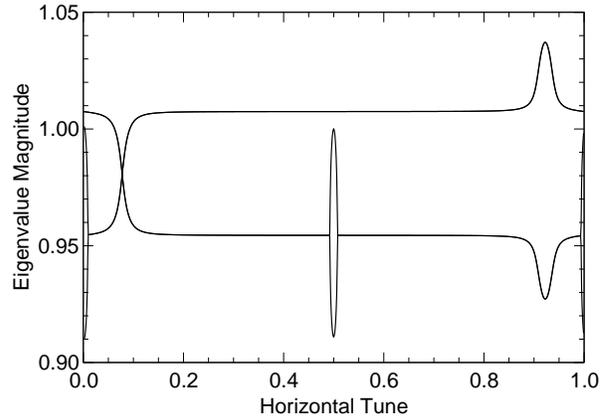}
  \caption{Magnitude of matrix eigenvalues as a function of
    $(\mu_++\mu_-)/(2\pi)$ for a lattice with dispersion at the RF cavities and
    an absorber with $\theta_1=\theta_2=0$.}
  \label{fig:reson}
\end{figure}
Begin with the case where there is dispersion in the RF cavity and
$\theta_0=\theta_1=0$.  Figure~\ref{fig:reson} shows the magnitude of
the eigenvalues as a function of $\mu_++\mu_-$, where I am taking
$\mu_+=\mu_-$.  Note that when $(\mu_++\mu_-)/(2\pi)\approx0.08$,
which is the synchrotron tune, then the magnitudes of all the
eigenvalues become less than 1~\cite{pac01:147}.
What one is seeing here is the
manifestation of a coupling resonance in a non-conservative dynamical
system.  Note that the same resonance phenomenon at
$(\mu_++\mu_-)/(2\pi)\approx1-0.08$ results in the longitudinal
eigenvalue becoming more unstable: instead of the resonance pulling
the eigenvalues together, it pushes them further apart.  The signs of
these effects are consistent with what one expects from the
perturbation calculation in Eq.~(\ref{eq:dispcav}): the sign of
$\sin(\mu_++\mu_-)$ gives the dominant effect for determining whether
the magnitude of the longitudinal eigenvalue increases or increases.
The loops in Fig.~\ref{fig:reson} near the tunes of 0 and 0.5 are not
important for this discussion:  they are simply the linear resonances of
the transverse lattice.

This example is furthermore an example of how a difference resonance
($\nu_x-\nu_z=k$ in this case) tends to lead to often innocuous
coupling between planes, whereas a sum resonance ($\nu_x+\nu_z=k$ in
this case) can lead to unbounded growth.  The coupling between the planes
is in fact an advantage in this case, since one is trying to make
the transverse damping affect the longitudinal plane.

Having a low horizontal tune is generally impractical for cooling channels,
since that tends to give a larger beta function at the absorber,
which results in a larger equilibrium emittance due to multiple
scattering~\cite{sjpn12:223,pa14:75}.
However, one could instead run in the passband with
horizontal tunes from 1 to 1.5, which would both create a low beta
function and ensure that coupling pulled the magnitudes of the eigenvalues
closer to each other.  It would also allow for a longer cell, making
it possible to increase the synchrotron tune, putting it further from
the integer resonance.  It may be more difficult to have a large momentum
acceptance in such a lattice, however, since the passband from tunes
of 1 to 1.5 generally would have a smaller relative momentum acceptance
than the lower passbands.

This example is meant more as a proof of principle, demonstrating that
one can in principle use any method of longitudinal-transverse coupling
to achieve cooling in all degrees of freedom.  The method may, however,
have practical applications in cases where it is impractical to control
the shape of the absorber.  One example might be the use of a lithium
lens in the final stages of cooling for a muon collider.  The method
may be more interesting at later stages of a muon collider in any
case, since the limited width of the resonance will likely translate
into a limited momentum acceptance for such a method, requiring a beam
which has already had its longitudinal emittance reduced significantly.

\begin{figure}
  \includegraphics[width=\linewidth]{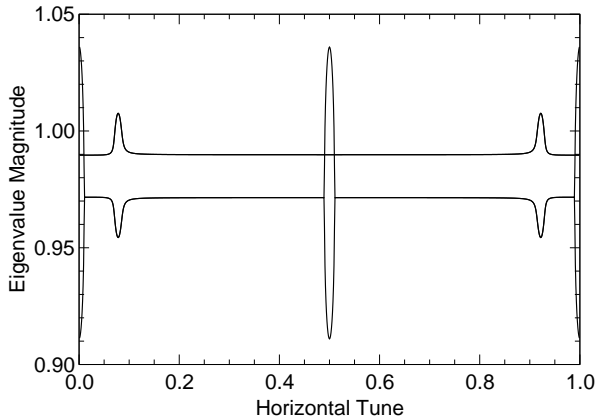}
  \caption{Magnitude of matrix eigenvalues as a function of
    $(\mu_++\mu_-)/(2\pi)$ for a lattice with dispersion at a wedge
    shaped absorber.}
  \label{fig:resonwedge}
\end{figure}
It is interesting to compare Fig.~\ref{fig:reson} with what one would
obtain when using a wedge-shaped absorber with dispersion.  This is
shown in Fig.~\ref{fig:resonwedge}.  Note that the eigenvalue
magnitudes are closer to each other over the entire range of
horizontal tunes (their values without dispersion at the wedges can be
deduced from Fig.~\ref{fig:reson}).  At both the sum and difference
resonances, the magnitudes are pushed further apart.  The difference
in the behavior near the synchro-betatron resonances
between this case and that shown in Fig.~\ref{fig:reson}
is a result of the non-symplectic nature of the coupling
that is generated by the wedge with dispersion.  Eqs.~(\ref{eq:fcoef})
and (\ref{eq:rotabs}) do an extremely good job of predicting the
behavior in Fig.~\ref{fig:resonwedge}, including the resonant behavior;
this is to be expected since the eigenvalues never get too close together,
even near the resonance.

\section{Physical Explanation}

To get some understanding of what is going on physically, one should
examine the eigenvectors.  The real and imaginary parts of the
eigenvectors define an ellipse that the particles move on.  Of course,
in the case of cooling, the radius of the particles are decreasing (or
sometimes increasing), but that piece can be taken out (i.e., the
difference of the magnitude of the eigenvalue from 1).  To see the
ellipse, take the real (or imaginary) part of the eigenvector,
multiply the vector by the matrix for the lattice cell, and divide the
result by the magnitude of the corresponding eigenvalue.  Repeat the
process, and one will trace out an ellipse in phase space.  Alternatively,
if $\vv$ is the eigenvector, then the ellipse is the set of points
obtained from
\begin{equation}
  \re\{\vv\}\cos u+\im\{\vv\}\sin u
  \label{eq:ellip}
\end{equation}
by varying $u$ from 0 to $2\pi$.

In the case where there is no coupling, the eigenvectors have
components either entirely in the horizontal plane, or entirely in the
longitudinal (time-energy) plane.  However, once one introduces any
type of longitudinal-transverse coupling, all of the eigenvectors will
have components in both planes.  This is the key to what allows the
reduction in transverse momentum spread that the absorber accomplishes
to reduce the amplitude of what would otherwise be longitudinal
motion.  The only effect that actually reduces the beam emittance
is the reduction of transverse momentum in the absorber.

\begin{figure}
  \includegraphics[width=\linewidth]{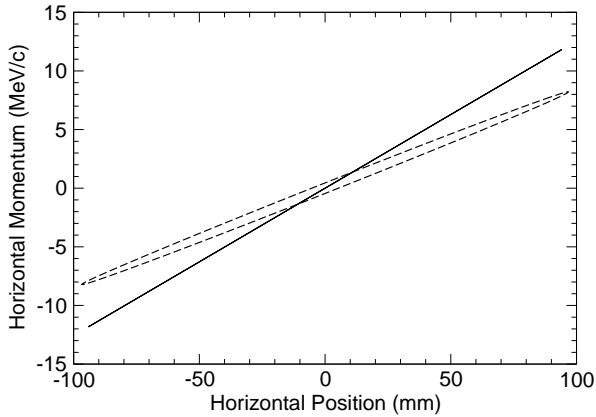}
  \caption{Projection of the longitudinal eigenvector at the absorber
    into the
    transverse plane when there is dispersion at the absorber.  The
    solid ellipse (which has no area, so it appears as a line) is for
    a parallel-faced absorber ($\theta_1=\theta_2=0$), whereas the
    dashed ellipse is for a wedge-shaped absorber.}
  \label{fig:ellipabs}
\end{figure}
Figure~\ref{fig:ellipabs} shows the projection of the
longitudinal eigenvector (identified by the phase of its corresponding
eigenvalue) at the absorber into the transverse plane.  The figure
shows two cases: one where there is dispersion at the absorber, but
the absorber has parallel faces with $\theta_1=\theta_2=0$, and one
where the absorber is wedge shaped, with $\theta_1=\theta_2>0$.  For
the parallel-faced case, there is no reduction in the magnitude of the
longitudinal eigenvalue, despite the fact that Fig.~\ref{fig:ellipabs}
shows that there is a nonzero projection of the longitudinal
eigenvalue onto the horizontal plane (and horizontal momentum in
particular).  The reason that there is no reduction in the magnitude
of the longitudinal eigenvalue is that the ellipse projected into
the horizontal plane has no area.  The absorber makes positive momenta
more negative and negative momenta more positive, reducing the
area of an ellipse.  But if the ellipse has no area, no reduction
can occur.  Thus, one needs the projected ellipse to have a nonzero
area to see an effect.  This is what making the absorber wedge-shaped
accomplishes, as can be seen in Fig.~\ref{fig:ellipabs}.

If one changes the horizontal tune to around 0.077, one can see from
Fig.~\ref{fig:resonwedge} that the longitudinal and transverse
eigenvalues return almost to their values without coupling.  This is
reflected in the eigenvectors in that the area of the horizontal
projection of the ellipse relative to the area of the longitudinal
projection is less than it was for horizontal tunes away from the longitudinal
tune (which was shown in Fig.~\ref{fig:ellipabs}).

\begin{figure}
  \includegraphics[width=\linewidth]{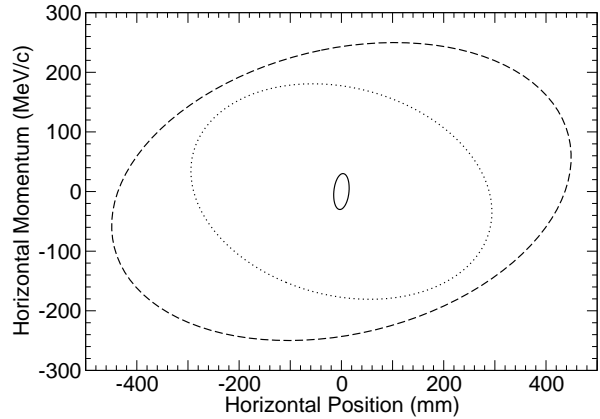}
  \caption{Projection of the longitudinal eigenvector at the absorber
    into the
    transverse plane when there is a parallel-faced absorber and
    dispersion only at the RF cavity.  The
    solid ellipse is for
    a horizontal tune of 0.75,
    the dotted ellipse for a horizontal tune of 0.09 (near
    one coupling resonance), and the dashed ellipse for a horizontal tune
    of 0.91 (near the other coupling resonance).}
  \label{fig:elliprf}
\end{figure}
Figure~\ref{fig:elliprf} shows the ellipses for the case where there
is dispersion at the RF cavity and not the absorber (this corresponds
to Fig.~\ref{fig:reson}).  The ellipses are smaller when one is away
from resonance, and larger when one is closer to resonance, reflecting
the stronger coupling between the eigenvalues near the resonance,
in particular that the magnitude of the longitudinal eigenvalue is
reduced (or increased for the higher-tune resonance) there.  What
the figure does not illustrate is why near the low tune resonance
(0.09 for the example in Fig.~\ref{fig:elliprf}), the magnitude of
the longitudinal eigenvalue is reduced, whereas near the high tune resonance,
the magnitude of the longitudinal eigenvalue is increased.

To understand this, one must first think about what area means for an ellipse
in four-dimensional phase space.  Hamilton's equations of motion can be
written as
\begin{equation}
  \dfrac{d\zv}{ds} = -J\nablav H(\zv,s),
\end{equation}
where for the example here where only horizontal and longitudinal
dimensions are considered,
\begin{equation}
  J =
  \begin{bmatrix}
    0&1&0&0\\-1&0&0&0\\0&0&0&-1\\0&0&1&0
  \end{bmatrix}.\label{eq:sympJ}
\end{equation}
One can define the area of a four-dimensional ellipse which is described
by Eq.~(\ref{eq:ellip}) to be $\pi$ times
\begin{equation}
  (\im\{\vv\})^TJ(\re\{\vv\}).
\end{equation}
This gives the expected answer when the ellipse lies entirely in either
the horizontal or longitudinal phase space plane, and is thus the
obvious extension of the concept.  In fact, it is really the sum
of the projected areas in the two planes.  Furthermore, it is
invariant under symplectic transformations of $\vv$, and thus reflects
the area-preserving nature of symplectic transformations.

Note that this area, as defined, has a sign.  Since the eigenvectors
come in complex-conjugate pairs, there are always two areas with the
same magnitude and opposite signs generated by the eigenvectors of a
matrix, assuming the eigenvectors were not normalized differently.
Furthermore, since the area is really a sum of the horizontal area
and the longitudinal area, these projected areas individually have
signs as well.  It turns out that when the longitudinal eigenvalue
gets closer in magnitude to the transverse one, the areas have the
same sign, whereas when they are pushed away from each other, the
areas have opposite signs.

Physically, this is because the sign of the ellipse area is related to
the direction which a particle moves around the ellipse in phase
space, as can be seen from Eq.~(\ref{eq:ellip}).  There is a
``physical'' direction that particles move around an ellipse, which is
reflected in the signs of the nonzero elements in $J$.  Normally,
particles with positive horizontal momentum will have an increasing
horizontal position; particles with a larger energy will have a
shorter time of flight (ignoring momentum compaction).  When the signs
of the projected areas (horizontal and longitudinal) are identical,
the particles are moving in the ``physical'' direction on both the
horizontal and longitudinal ellipses.  If, on the other hand, the
projected areas have opposite signs, then longitudinal motion
in the ``physical'' direction will appear to have ``unphysical''
motion in the horizontal plane.  Thus, in particular, the reduction
in horizontal momentum that the absorber accomplishes will not
be translated physically into the area reduction that one hopes for,
since the change in horizontal position goes the wrong way.  The result
is the eigenvalue magnitudes getting further apart, as seen
on the right hand side of Fig.~\ref{fig:reson}.

\section{Conclusion}

I have shown that one can obtain ionization cooling in all phase
space planes through two different methods: by generating dispersion
at the absorber and shaping the absorber appropriately (this is the usual
method), and by
using the longitudinal-transverse coupling resonance with a
parallel-faced absorber.  I have approximated the effect 
by perturbation theory, and using that have demonstrated that 
one should shape the absorber in conformity to the
dispersion at the absorber,
in particular taking into account whether the dispersion
is in position or transverse momentum.  I have demonstrated that it
is the fact that the longitudinal motion has a component in the
transverse plane that allows the absorber to cause cooling in the
longitudinal plane as well as the transverse.  The strength
of the effect is related to the area, including sign, of the ``longitudinal''
ellipse
projected into the horizontal plane.

\bibliographystyle{elsart-num}
\bibliography{060701}
\end{document}